\documentclass[pra,amsmath,amssymb,twocolumn,superscriptaddress,floatfix]{revtex4-2}
\usepackage[english]{babel}
\usepackage{amsmath,amssymb}
\usepackage{grffile}
\usepackage{textcomp}
\usepackage{xspace}
\usepackage{graphicx} %
\usepackage{natbib}
\usepackage[dvipsnames]{xcolor}
\usepackage{ulem}
\usepackage{dsfont} 
\usepackage[acronym]{glossaries} 
\usepackage{multirow} 
\usepackage{placeins} 
\usepackage[qm]{qcircuit} 
\usepackage{floatrow}
\usepackage{comment} 
\usepackage{braket}
\usepackage{bbold}
\usepackage{verbatim}
\usepackage{tikz}
\usepackage{fontenc}
\usepackage[caption=false]{subfig}
\newcommand{\mydarkblue}{Blue} 
\usepackage[colorlinks, citecolor=black, linkcolor=black, urlcolor=\mydarkblue, breaklinks]{hyperref}

\newcommand{\im}{{\rm i}}

\newcommand{\sx}{\hat{\sigma}^x}

\newcommand{\sz}{\hat{\sigma}^z}

\newcommand{\hsb}{\hat{H}_{SB}}
\newcommand{\parasf}{$\epsilon=0.5$, $\omega=4$, $\lambda=2$}
\newcommand{\parasff}{$\epsilon=0.5$, $\omega=6$, $\lambda=2$}
\newcommand{\dt}{\Delta t}

\newcommand{\sumindex}{k}
\newcommand{\ident}{\mathds{1}}
\newcommand{\fid}{\mathcal{F}}
\newcommand{\ifid}{\mathcal{I}}

\newcommand{\expect}[1]{\langle #1 \rangle } 
\newcommand{\pz}{\hat{\sigma}^{z}}
\newcommand{\px}{\hat{\sigma}^{x}}
\newcommand{\py}{\hat{\sigma}^{y}}
\newcommand{\pmi}{\hat{\sigma}^{-}}
\newcommand{\ppl}{\hat{\sigma}^{+}}
\newcommand{\opa}{\hat{a}}
\newcommand{\opad}{\hat{a}^{\dagger}}
\newcommand{\ada}{\hat{a}^{\dagger}\hat{a}}

\newcommand{\opl}{\hat{L}_{\sumindex}}
\newcommand{\opld}{\hat{L}_{\sumindex}^{\dagger}}

\newcommand{\cxgate}{CX-Gate}
\newcommand{\xgate}{X-Gate}

\newcommand{\dho}{d_{HO}}

\newcommand{\szcorr}{C^{ZZ}}
\newcommand{\sxcorr}{C^{XX}}
\newcommand{\szcorrdef}{\langle\sz_1\sz_2\rangle - \langle\sz_1\rangle \langle\sz_2\rangle}
\newcommand{\sxcorrdef}{\langle\sx_1\sx_2\rangle - \langle\sx_1\rangle \langle\sx_2\rangle}

\newcommand{\mult}{\cdot} 

\newcommand{\dm}{\hat{\rho}}
\newcommand{\probp}{\mathcal{P}} 
\newcommand{\channelthermal}{\mathcal{E}_T}
\newcommand{\channeldepol}{\mathcal{E}_D}
\newcommand{\temperature}{\mathcal{T}}  
\newcommand{\tOne}{T_1}
\newcommand{\tTwo}{T_2}
\newcommand{\ifidGate}{\ifid_{Gate}}
\newcommand{\ifidDepol}{\ifid_{D}}
\newcommand{\ifidThermal}{\ifid_{T}}
\newcommand{\freqQubit}{f_{Qubit}}
\newcommand{\timeGate}{t_{Gate}}
\newcommand{\tempQubit}{\temperature_{Qubit}}
\newcommand{\dmapprox}{\dm^{\prime}}  
\newcommand{\numspins}{N_{S}}

\newcommand{\sutd}{Science, Mathematics and Technology Cluster, Singapore University of Technology and Design, 8 Somapah Road, 487372 Singapore}
\newcommand{\sutdepd}{EPD Pillar, Singapore University of Technology and Design, 8 Somapah Road, 487372 Singapore}
\newcommand{\cqt}{Centre for Quantum Technologies, National University of Singapore, 117543 Singapore}
\newcommand{\nie}{National Institute of Education and Institute of Advanced Studies, Nanyang Technological University 637616, Singapore}  
\newcommand{\majulab}{MajuLab, CNRS-UNS-NUS-NTU International Joint Research Unit UMI 3654, Singapore} 
\newcommand{\munich}{Faculty of Physics, Ludwig-Maximilians-Universit\"at Munich,
\\Geschwister-Scholl-Platz 1, 80539 Munich, Germany} 

\begin{document}
\title{
Digital Quantum Simulation of the Spin-Boson Model \\ under 
Markovian
Open System Dynamics
}

\author{Andreas Burger} 
\affiliation{\munich}
\affiliation{\sutd} 
\affiliation{\cqt}
\author{Leong Chuan Kwek}
\affiliation{\cqt}
\affiliation{\nie} 
\affiliation{\majulab}
\author{Dario Poletti} 
\affiliation{\sutd} 
\affiliation{\sutdepd} 
\affiliation{The Abdus Salam International Centre for Theoretical Physics, Strada Costiera 11, 34151 Trieste, Italy} 
\affiliation{\cqt} 
\affiliation{\majulab}

\begin{abstract}
Digital quantum computers have the potential to simulate complex quantum systems. The spin-boson model is one of such systems, used in disparate physical domains. Importantly, in a number of setups, the spin-boson model is open, i.e. the system is in contact with an external environment which can, for instance, cause the decay of the spin state. Here we study how to simulate such open quantum dynamics in a digital quantum computer, for which we use one of IBM's hardware. We consider in particular how accurate different implementations of the evolution result as a function of the level of noise in the hardware and of the parameters of the open dynamics. For the regimes studied, we show that the key aspect is to simulate the unitary portion of the dynamics, while the dissipative part can lead to a more noise-resistant simulation. We consider both a single spin coupled to a harmonic oscillator, and also two spins coupled to the oscillator. In the latter case, we show that it is possible to simulate the emergence of correlations between the spins via the oscillator.   
\end{abstract}
  
\maketitle

\setcounter{figure}{0}

\section{Introduction}


A natural application of quantum computers is the simulation of quantum systems \cite{Lloyd1996, Nielsen2010}.  And most hardware realizations of quantum computers implement the qubit.  A prevalent qubit-based quantum system is the spin system. Existing quantum computers are based on unitary quantum circuits. Consequently, there has been a plethora of research on closed quantum systems. 
\cite{Whitfield2011, Wiebe_2011, Tacchino2019, Jaderberg2022}. 
Amongst the spin models, an important class is the spin-boson problem, where one or more spins are coupled to several bosonic degrees of freedom. These models possess rich many-body physics and they can model realistic coupling between electron transfer and protein motion or a solvent \cite{Leggett1987, Weiss2011, Xu1994, Renger2002, Fleming1996}.



In the last few years, NISQ computers \cite{Preskill2018, Bharti2022} have offered a new perspective on the
implementations on digital devices, leading to an explosion of activities.
Not all computing tasks are amenable to quantum processing. Classical optimization can often perform better than quantum algorithms. 
The challenges of device-induced noise have led to the popularity of hybrid quantum-classical variational algorithms (VQA) that split the workload between a quantum and a classical processor. These techniques are ideally  suited for  the evaluation of different quantities such as eigenstates \cite{Peruzzo2014}, general quantum approximate optimization algorithms \cite{Farhi2014}, off-diagonal elements of matrices \cite{Erbanni2022} and more \cite{Bharti2022}. Importantly, new error mitigation approaches have also been proposed \cite{Li2017, Temme2017, Endo2021}.        
VQA has been applied to boson-spin systems or its equivalents \cite{DiPaolo2020, Miessen2021, Fitzpatrick2021}. 
Regarding open systems, different VQA approaches have been tested. They include approaches based on imaginary time evolution \cite{Kamakari2022, McArdle2019},
stochastic Schr\"odinger equation \cite{Endo2020}, variational quantum eigensolvers to reach steady states \cite{Fujii2020, Liu2021},
and the quantum assisted simulator without a classical-quantum feedback loop \cite{Bharti2021}. 
Mapping bosonic problems to quantum circuits has been laid out in \cite{Macridin2018, Somma2003, Sawaya2020}, while a recent implementation of spin-boson models can be found in \cite{Jaderberg2022}. 


Simulating open quantum systems entirely on digital quantum computers has primarily focused around two-level systems.
The amplitude damping channel has been implemented with a unitary dilation of the Kraus operators \cite{Hu2020},
using uniformly controlled gates \cite{Schlimgen2021, Udayakumar2019}, 
and with the amplitude damping circuit \cite{Nielsen2010, GarciaPerez2020}. 
Larger systems have been realized using linear combination of unitary matrices \cite{Wei2016, Cleve2019} and modified stochastic Schr\"odinger equation methods \cite{Jo2022}. In \cite{Endo2020}, the authors proposed a hybrid classical-quantum variational approach to simulate generic Markovian open quantum systems.   


\begin{figure}
\includegraphics[width=0.8\columnwidth]{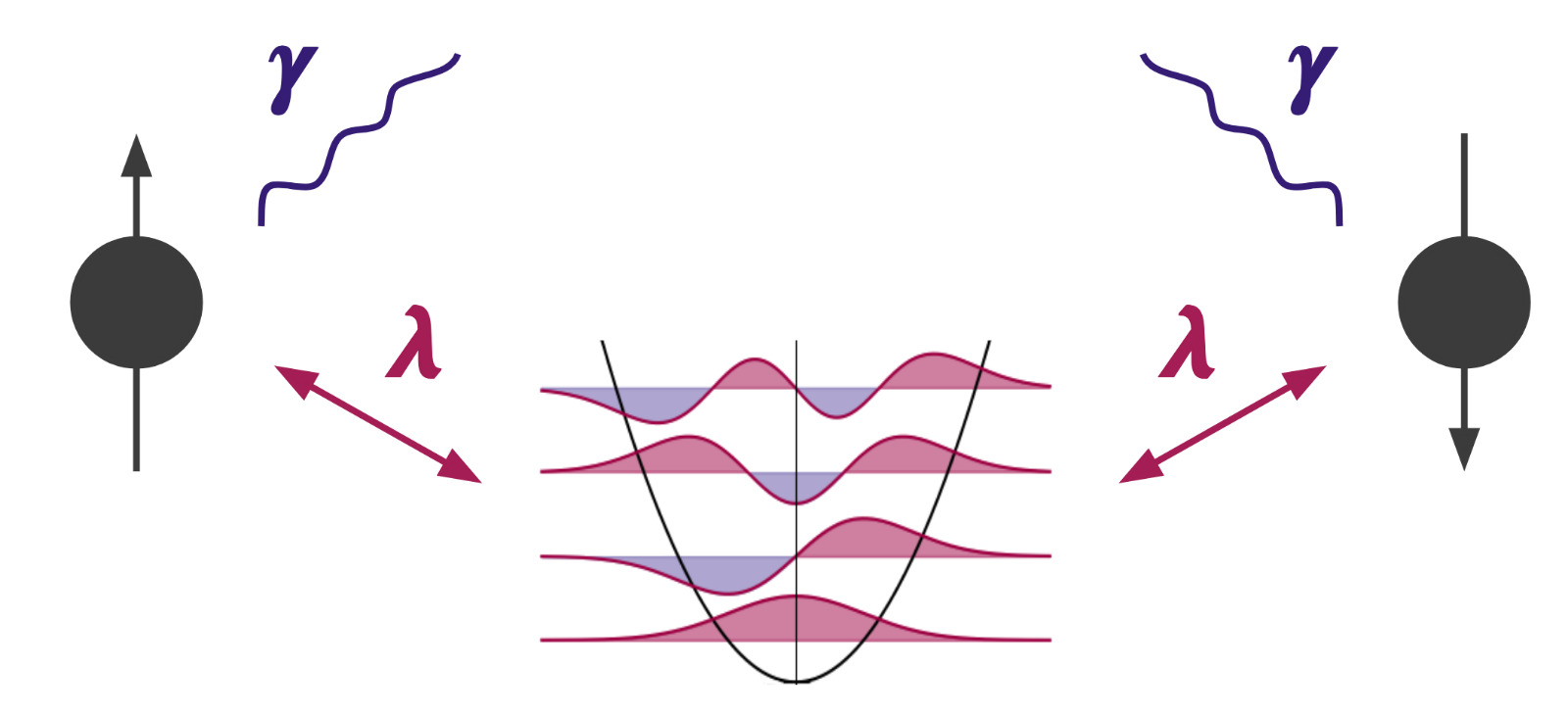}
\caption{Depiction of model described by Eqs.(\ref{eq:spinboson},\ref{eq:gksl}) for a number of spins $N_S=2$. The two spin sites are coupled to one harmonic oscillator of frequency $\omega$ via coupling parameter $\lambda$. Each of the spins dissipates independently into the environment at a rate $\gamma$.}
\label{fig:sketch_model}
\end{figure}

Our aim is to simulate the open dynamics of a spin-boson model coupled to a dissipative channel on a digital quantum computer. 
We do this by mapping the bosonic modes to qubits, Trotterizing the unitary evolution, and modeling the dissipative portion via repeated collisions with a resetted auxiliary qubit \cite{Ciccarello2022, GarciaPerez2020, Algaba2022}. 
In doing so we focus on using different noise levels in the quantum computer, from the value in current hardware, to $1\%$ of it. 
With this in mind we study how different implementations of the simulation perform in presence of different noise levels. 

The paper is organized as follows.
In Sec.~\ref{sec:spin_boson_model},
we introduce the open spin-boson model 
and lay out the circuit implementation.
In Sec.~\ref{sec:circuit_implementation}, we describe the circuit implementation of the unitary and dissipative evolutions. 
We then detail our use of quantum hardware and noise-related limitations of the devices \ref{sec:hardware}.
Our results are presented in Sec.~\ref{sec:results}. 
We quantify the error stemming from approximations in the model, and for different magnitudes of noise in the device. 
We study the optimal time-step-sizes and dissipative rates in terms of fidelity.
Finally, we increase the system size to two spins and investigate if it is possible to observe rising correlations amongst the spins.


\section{Method\label{sec:method}}

\subsection{Model\label{sec:spin_boson_model}}

We consider $N_S$ non-interacting spins coupled to a single harmonic oscillator, as well as to a bath, see Fig.~\ref{fig:sketch_model}. 
The closed system is governed 
by the quantum Rabi Hamiltonian \cite{Rabi1936, Rabi1937, Bloch1940}, which describes the ultra-strong coupling regime, where the usual rotating wave approximation breaks down and the counter-rotating term can no longer be neglected \cite{Jaynes1963, Cummings2013, Xie2017}.
\begin{equation} \label{eq:spinboson}
\hsb = \hbar\omega \ada +
\sum_{i=1}^{N_S}  
\frac{1}{2} (h\pz_{\sumindex} 
+ \epsilon \px_{\sumindex} )
+
\lambda \px_{\sumindex} \ (\opad + \opa),  
\end{equation} 
Experimentally the ultra-strong coupling regime has been investigated in circuit QED \cite{FornDiaz2010, Niemczyk2010, Braumueller2017, FornDiaz2016, Yoshihara2016, Langford2017}, trapped ions \cite{Lv2018}, photonic systems \cite{Crespi2012} and semiconductors \cite{Todorov2009, Guenter2009}.

In Eq.~(\ref{eq:spinboson}), $\opad$ and $\opa$ respectively create and destroy one excitation in the harmonic oscillator while $\px_{\sumindex}=\ppl_{\sumindex}+\pmi_{\sumindex}$ and $\pz_{\sumindex}$ are Pauli operators acting on the spin(s). 
$h$ is the local magnetic field in the $z-$direction while $\epsilon$ is a field in the $x-$direction. $\lambda$ is the magnitude of the coupling between the spins and the harmonic oscillator, with frequency $\omega$. 
In the following we will work in units such that $h=\hbar=1$. 

The dissipative part of the dynamics is here described by a Markovian master equation in Gorini-Kossakovski-Sudarshan-Lindblad form \cite{Lindblad1976, Gorini1976} 
\begin{align} \label{eq:gksl}
    \frac{d\dm}{dt} &= -\frac{\im}{\hbar} [\hsb,\dm] + \gamma \sum_{\sumindex}( 2 \opl \dm \opld - \{\opld \opl, \dm\}) 
\end{align}
with the amplitude damping channel  $\opl = \ket{\downarrow}_{\sumindex} \bra{\uparrow} $ acting on the $\sumindex-$th spin and $\gamma$ being the decay rate. 
$\ket{\downarrow}$ represents the vacuum state, whereas $\ket{\uparrow}$ represents the excited state of the spin.

Eq.~(\ref{eq:gksl}) describes a setup where loss from imperfections in the cavity are negligible compared to the spins emissions.
In these systems undesired decay transitions can include emission of frequencies which are suppressed in the cavity and thus are effectively lost \cite{Ritsch2013, Reiserer2015, Fabre2017}.

\subsection{Circuit implementation}\label{sec:circuit_implementation}
In this section we describe how we implement the evolution governed by Eqs.~(\ref{eq:spinboson},\ref{eq:gksl}) in a quantum circuit. 

\paragraph{Encoding of the Hamiltonian\label{sec:encoding}}

We map the spin and bosonic operators in $\hsb$ to Pauli operators, and Trotterize the unitary $e^{-i\hsb t}$.
The spin part is trivially mapped to qubits. 
For the bosonic subspace and operators, we use a d-level-to-qubit mapping with Gray Code as the integer-to-bit encoding, as described in \cite{DiMatteo2021, Sawaya2020}.  
We have given more details of the mapping to $Q_B$ qubits in Appendix \ref{sec:boson_to_qubit}.

\paragraph{Trotterization of unitary\label{sec:trotter}}
To implement the unitary evolution operator $U= e^{-i \hsb t}$ we consider the first-order $U_1$ and second-order $U_2$ Suzuki-Trotter product formulas \cite{Hatano2005, Berry2006}
\begin{align}
    &U_1  = (e^{-ih_1 \Delta t }\ e^{-ih_2 \Delta t }\ ...\ e^{-ih_N \Delta t })^{ \frac{t}{\Delta t}} \label{eq:firstorder}
    \\ &U_2=
    (e^{-ih_1 \frac{\Delta t}{2} }
    \  ...
    \ e^{-ih_N \frac{\Delta t}{2}} 
    \ e^{-ih_N \frac{\Delta t}{2}} 
    \ ...
    \ e^{-ih_1 \frac{\Delta t}{2} } 
    )^{ \frac{t}{\Delta t}}  \label{eq:secondorder}
\end{align}
where $h_{\sumindex}$ are $N$ different, non-commuting, terms of the Hamiltonian after encoding and $\Delta t = t/N$. The individual exponentials of Pauli strings $e^{-ih_{\sumindex} \Delta t } $ are then implemented via the CNOT-staircase  \cite{Nielsen2010, Whitfield2011}, which is taken care of by Qiskit \cite{qiskit}. 
See Eqs.~(\ref{eq:hsb_digit_1}, \ref{eq:hsb_digit_2}) in Appendix \ref{sec:boson_to_qubit} for more details on $h_{\sumindex}$. 

\paragraph{Collisional model\label{sec:Collision}}
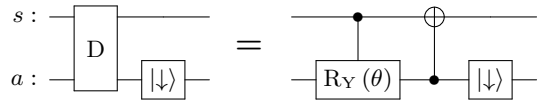
\begin{figure}
\subfloat{%
\scalebox{1.0}{ 
\scalebox{1.0}{
\Qcircuit @C=1.0em @R=1.0em @!R { \\
	 	\nghost{{s} :  } & \lstick{{s} :  } & \multigate{1}{\mathrm{D}}_<<<{} & \qw & \qw 
   \\
	 	\nghost{{a} :  } & \lstick{{a} :  } & \ghost{\mathrm{D}}_<<<{} & \gate{\mathrm{\ket{\downarrow}}} & \qw 
   \\
\\ }
}
\scalebox{1.0}{
\centering
\Qcircuit @C=0.5em @R=1.0em @!R { \\
	 	\nghost{{s}} & \lstick{}  \\
	 	\nghost{{a}} & \lstick{ \text{\large \textbf{=}} } \\ 
\\ }
}
\scalebox{1.0}{
\Qcircuit @C=1.0em @R=1.0em @!R { \\
& \ctrl{1} & \targ 
& \qw & \qw\\
& \gate{\mathrm{R_Y}\,(\mathrm{\theta})} & \ctrl{-1} & \gate{\mathrm{\ket{\downarrow}}} & \qw 
\\
\\ }
}
} 
} 
\caption{Circuit implementation of the dissipative part of the circuit, $D$, which represent a single collision to model Eq.~(\ref{eq:gksl}). $s$ is the qubit representing the spin while $a$ represents the auxiliary qubit.   
\label{fig:adc_circuit}
}
\end{figure}
We model the local master equation Eq.~(\ref{eq:gksl}) via repeated collisions \cite{Ciccarello2022, Karevski2009}.  
Fig.~\ref{fig:adc_circuit} gives a depiction of a single collision. We consider the spin qubit $s$, and auxiliary qubit $a$ and where a controlled-$R_Y(\theta)$ (rotation around y-axis) is followed by a controlled-NOT and a reset of the auxiliary qubit, see Appendices \ref{sec:gate_definition} and \ref{sec:transpiled} for more details. To reproduce Eq.~(\ref{eq:gksl}) we use $\theta =  \arcsin \left(\sqrt{1-e^{-\gamma t}} \right)$ \cite{Nielsen2010}. 

\paragraph{Integration of dissipative and unitary part} 

To integrate the step of Fig.~\ref{fig:adc_circuit} in the main circuit, we employ a first-order Suzuki-Trotter decomposition which alternates between the unitary and the dissipative parts. In Fig.~\ref{fig:Collision_circuit}(a) we depict three steps of the evolution of a single spin coupled to a harmonic oscillator mapped to two qubits, while in Fig.~\ref{fig:Collision_circuit}(b) we show our implementation of three-step evolution of the case with two spins and one harmonic oscillator. For considerations of connectivity, the auxiliary qubits needed for the dissipative channel are placed at the edges of the circuit, next to the spins. 
After all time-steps are finished, the qubits representing the spin(s) $s_{\sumindex}$ and the bosons $b_{\sumindex}$ are measured, while the state of the auxiliary qubit is ignored.

\newcommand{\measuregate}{\meter}  
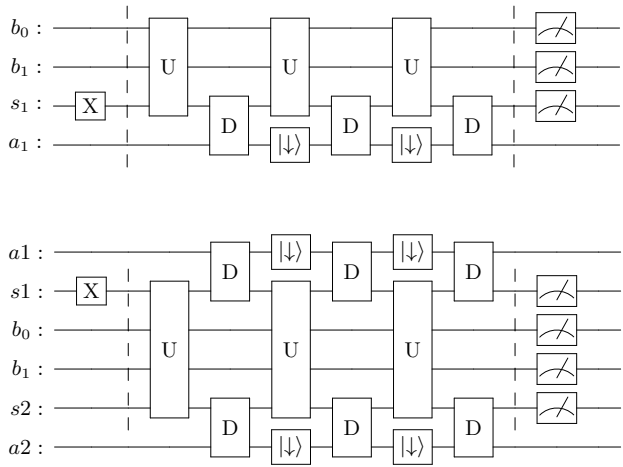
\begin{figure}
\subfloat{%
\scalebox{0.9}{
\Qcircuit @C=1.0em @R=0.2em @!R { \\
 \nghost{{b}_{0} :  } & \lstick{{b}_{0} :  } & \qw \barrier[0em]{3} & \qw & \multigate{2}{\mathrm{U}}_<<<{} & \qw & \multigate{2}{\mathrm{U}}_<<<{} & \qw & \multigate{2}{\mathrm{U}}_<<<{} & \qw \barrier[0em]{3} & \qw & \measuregate & \qw & \qw\\
 \nghost{{b}_{1} :  } & \lstick{{b}_{1} :  } & \qw & \qw & \ghost{\mathrm{U}}_<<<{} & \qw & \ghost{\mathrm{U}}_<<<{} & \qw & \ghost{\mathrm{U}}_<<<{} & \qw & \qw & \measuregate & \qw & \qw\\
 \nghost{{s_1} :  } & \lstick{{s_1} :  } & \gate{\mathrm{X}} & \qw & \ghost{\mathrm{U}}_<<<{} & \multigate{1}{\mathrm{D}}_<<<{} & \ghost{\mathrm{U}}_<<<{} & \multigate{1}{\mathrm{D}}_<<<{} & \ghost{\mathrm{U}}_<<<{} & \multigate{1}{\mathrm{D}}_<<<{} & \qw & \measuregate & \qw & \qw\\
 \nghost{{a_1} :  } & \lstick{{a_1} :  } & \qw & \qw & \qw & \ghost{\mathrm{D}}_<<<{} & \gate{\mathrm{\ket{\downarrow}}} & \ghost{\mathrm{D}}_<<<{} & \gate{\mathrm{\ket{\downarrow}}} & \ghost{\mathrm{D}}_<<<{} & \qw & \qw & \qw & \qw\\
\\ }
}
} 

\subfloat{%
\scalebox{0.9}{
\Qcircuit @C=1.0em @R=0.2em @!R { \\
 \nghost{{a1} :  } & \lstick{{a1} :  } & \qw & \qw & \qw & \multigate{1}{\mathrm{D}}_<<<{} & \gate{\mathrm{\ket{\downarrow}}} & \multigate{1}{\mathrm{D}}_<<<{} & \gate{\mathrm{\ket{\downarrow}}} & \multigate{1}{\mathrm{D}}_<<<{} & \qw & \qw & \qw & \qw\\
 \nghost{{s1} :  } & \lstick{{s1} :  } & \gate{\mathrm{X}} \barrier[0em]{3} & \qw & \multigate{3}{\mathrm{U}}_<<<{} & \ghost{\mathrm{D}}_<<<{} & \multigate{3}{\mathrm{U}}_<<<{} & \ghost{\mathrm{D}}_<<<{} & \multigate{3}{\mathrm{U}}_<<<{} & \ghost{\mathrm{D}}_<<<{} \barrier[0em]{3} & \qw & \measuregate & \qw & \qw\\
 \nghost{{b}_{0} :  } & \lstick{{b}_{0} :  } & \qw & \qw & \ghost{\mathrm{U}}_<<<{} & \qw & \ghost{\mathrm{U}}_<<<{} & \qw & \ghost{\mathrm{U}}_<<<{} & \qw & \qw & \measuregate & \qw & \qw\\
 \nghost{{b}_{1} :  } & \lstick{{b}_{1} :  } & \qw & \qw & \ghost{\mathrm{U}}_<<<{} & \qw & \ghost{\mathrm{U}}_<<<{} & \qw & \ghost{\mathrm{U}}_<<<{} & \qw & \qw & \measuregate & \qw & \qw\\
 \nghost{{s2} :  } & \lstick{{s2} :  } & \qw & \qw & \ghost{\mathrm{U}}_<<<{} & \multigate{1}{\mathrm{D}}_<<<{} & \ghost{\mathrm{U}}_<<<{} & \multigate{1}{\mathrm{D}}_<<<{} & \ghost{\mathrm{U}}_<<<{} & \multigate{1}{\mathrm{D}}_<<<{} & \qw & \measuregate & \qw & \qw\\
 \nghost{{a2} :  } & \lstick{{a2} :  } & \qw & \qw & \qw & \ghost{\mathrm{D}}_<<<{} & \gate{\mathrm{\ket{\downarrow}}} & \ghost{\mathrm{D}}_<<<{} & \gate{\mathrm{\ket{\downarrow}}} & \ghost{\mathrm{D}}_<<<{} & \qw & \qw & \qw & \qw\\
\\ }
}   
} 
\caption{Circuit structure, alternating between a unitary evolution and collisions with auxiliary qubits and resets. Here the spins are represented by $s_{\sumindex}$, the harmonic oscillator modes (4 levels) are encoded on the $b_{\sumindex}$ qubits and the auxiliary qubits are represented by $a_{\sumindex}$.
\xgate{} represents the initial state preparation, $\ket{\downarrow}$ represent resets, the final gates represent measurements, while $D$ is described in Fig.~\ref{fig:adc_circuit}. 
}
\label{fig:Collision_circuit}
\end{figure}

\subsection{Quantum hardware simulation}\label{sec:hardware}
To perform our quantum circuit simulations and run it on actual quantum hardware, we use IBM's Qiskit software \cite{qiskit}.
The Quantum Computer we use is the 7-qubit ibmq\_jakarta device with a native gate set \{CNOT, ID, RZ, SX, X\}. 
Each circuit is run with $2^{13} = 8192$ shots (repetitions).

We quantify the error at each point in time as the infidelity $\ifid$ \cite{Jozsa1994}
\begin{equation}
\ifid(\dm, \dmapprox) = 
1- \left( \text{Tr} \left[ \sqrt{\sqrt{\dm} \dmapprox \sqrt{\dm}} \right] \right)^2 
\end{equation}
where we obtain the density matrix $\dmapprox$ of the circuit via quantum state tomography. 
We also consider a time-averaged version of the infidelity $\bar{\ifid}$, which is obtained by averaging the infidelity over time, except for the time $t=0$ which consists of just the state preparation. The exact density matrix $\dm$ for the benchmark is obtained from exact evolution of the master equation (Eq.~\ref{eq:gksl}), for which we use QuTiP \cite{qutip}. 
To mitigate the measurement error on noisy hardware, we classically post-process the results with Qiskit's error mitigation, which approximates the inverse of the noise matrix of the readout \cite{Bravyi2021}. 

\paragraph{Reduced-Noise Models} 
While it is important to study how current quantum processors can evaluate the model we study, we also aim to explore what could be the performance of future, less noisy, hardware. 
To model these scenarios, we use the same error channels that IBM uses to describe their current devices. 

The noise models include error sources in the gates, as thermal relaxation (relaxation and dephasing) and depolarizing errors, and also readout errors \cite{Georgopoulos2021}. 

For our reduced-noise models we scale down the average gate infidelity $\ifidGate$, the gate times $\timeGate$, and the false-readout probabilities, probability of measuring $1$ when the state is $0$ $P(1|0)$ or vice versa $P(0|1)$, by the same noise-factor $\xi$, or more precisely 
%
\begin{align}
    \ifidGate &\rightarrow \xi \mult \ifidGate \\
    \timeGate &\rightarrow \xi \mult \timeGate \\
    P(1|0),\ P(0|1) &\rightarrow \xi \mult P(1|0),\ \xi \mult P(0|1) 
\end{align} 
where $\xi$ ranges from $0$ to $1$.

Indeed, realistically some of these parameters will not see equal improvement in the next years, but a more detailed analysis of differentiated improvements of different aspects is beyond the scope of this work. 
%
Details on the error channels can be found in Appendix \ref{sec:noise_model}.


\section{Results\label{sec:results}}

Inaccuracies of the implementation of the model on a quantum computer can stem from different causes of completely different nature. We will first consider errors that rise from the Trotterization of the evolution in Sec.~\ref{sec:implementation_error}. We will then consider errors due to the noisy nature of the quantum computer in Sec.~\ref{sec:device_error}. In Sec.~\ref{sec:twospins} we will then study the case of two spins coupled to the harmonic oscillator.    

In the following, for the Hamiltonian, we choose the parameters \parasf{} for one spin, and \parasff{} for two spins. 
For the open dissipative rate we choose $\gamma=1$. With these parameters, an accurate evolution of the system up to a time $t=2$ can be obtained considering simply four levels for the harmonic oscillator, which can then be encoded with two qubits.  
For the initial state, we consider a pure product state between spins and bosons, with one spin in the excited state and zero excitations in the harmonic oscillator.
This choice of initial conditions allows observing oscillatory, non-trivial dynamics from early times, while not requiring too many levels for the harmonic oscillator.

%
\subsection{Error from the circuit implementation \label{sec:implementation_error}}

\begin{figure}
\centering
\subfloat{%
\centering
\includegraphics[width=1\columnwidth]{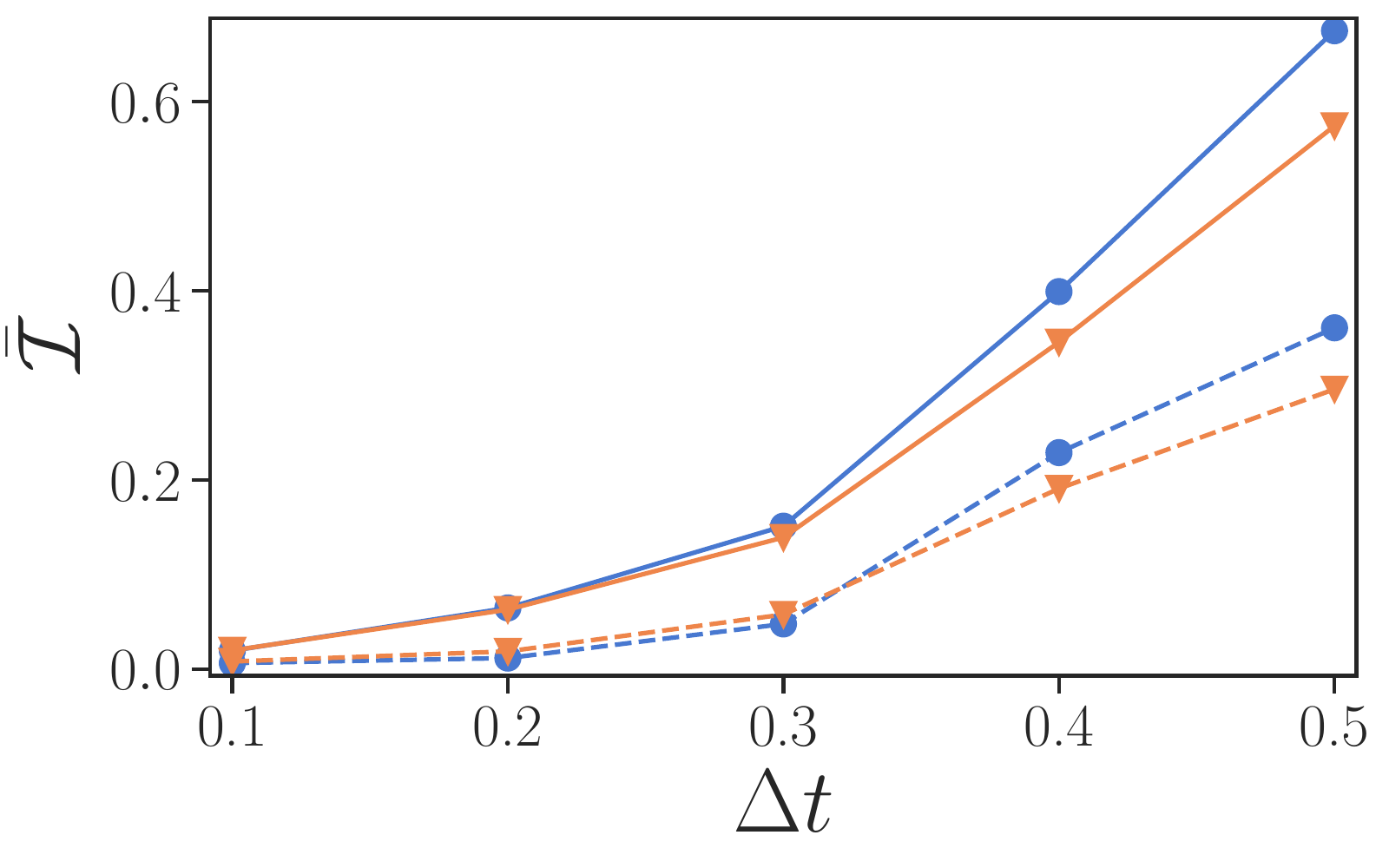}
}
\caption{Time-averaged infidelity for the evolution from $t=0$ to $t=2$. 
Noiseless simulations of the Hamiltonian $\gamma=0$ (blue line, dots) and the open system $\gamma=1$ (orange line, crosses).
The solid and dashed lines are used respectively for first-order and second-order Trotter implementations. The common parameters are \parasf.
The number of time-steps for $\Delta t=t/N=0.1,0.2,0.3,0.4,0.5$ are $N=20,10,7,5,4$ respectively, not counting $t=0$, which just consists of the initial state preparation.
}
\label{fig:error_model}
\end{figure}

As explained earlier, to implement the open dynamics, we Trotterize the unitary and dissipative parts of the master equation. 
However, also for the implementation of the unitary evolution, we need to rely on another layer of Trotterization. 
In Fig.~\ref{fig:error_model} we consider a unitary evolution with Hamiltonian $\hsb$ from Eq.~(\ref{eq:spinboson}) for a time-step $\dt$ and the possible implementation error, but considering no noise from the machine (blue lines). 
Implementing the various non-commuting terms of $\hsb$ in Qiskit \cite{qiskit} requires $48$ single-qubit- and $19$ \cxgate s or $79$ single-qubit- and $28$ \cxgate s, when using first or second-order Trotter respectively (Tab. \ref{tab:encoding_depth}).

In Fig.~\ref{fig:error_model} we evaluate the infidelity both for unitary and dissipative evolutions, i.e. following Eq.~(\ref{eq:gksl}) for $\gamma=0$ (blue lines with circles) or $\gamma=1$ (orange lines with triangles), versus $\Delta t$.  
We observe that the second-order Trotterization, dashed lines, has significantly smaller infidelity than a first-order implementation, continuous lines. 
Interestingly, beyond $\dt \approx 0.3$, the infidelity in just the Hamiltonian simulation is larger than the infidelity when including the dissipation.
Furthermore, independently on whether one considers first-order or second-order Trotterization, the dissipative dynamics has either smaller infidelity or it is very close to the unitary case.  
This implies that the unitary step 
implementing the Hamiltonian
is the main contribution to the infidelity 
compared to the implementation of the dissipation.

\subsection{Error in presence of noise \label{sec:device_error}} 
\captionsetup[subfigure]{labelformat=empty}

\begin{figure}[h]
\centering
\subfloat{
\centering
\includegraphics[width=\columnwidth]{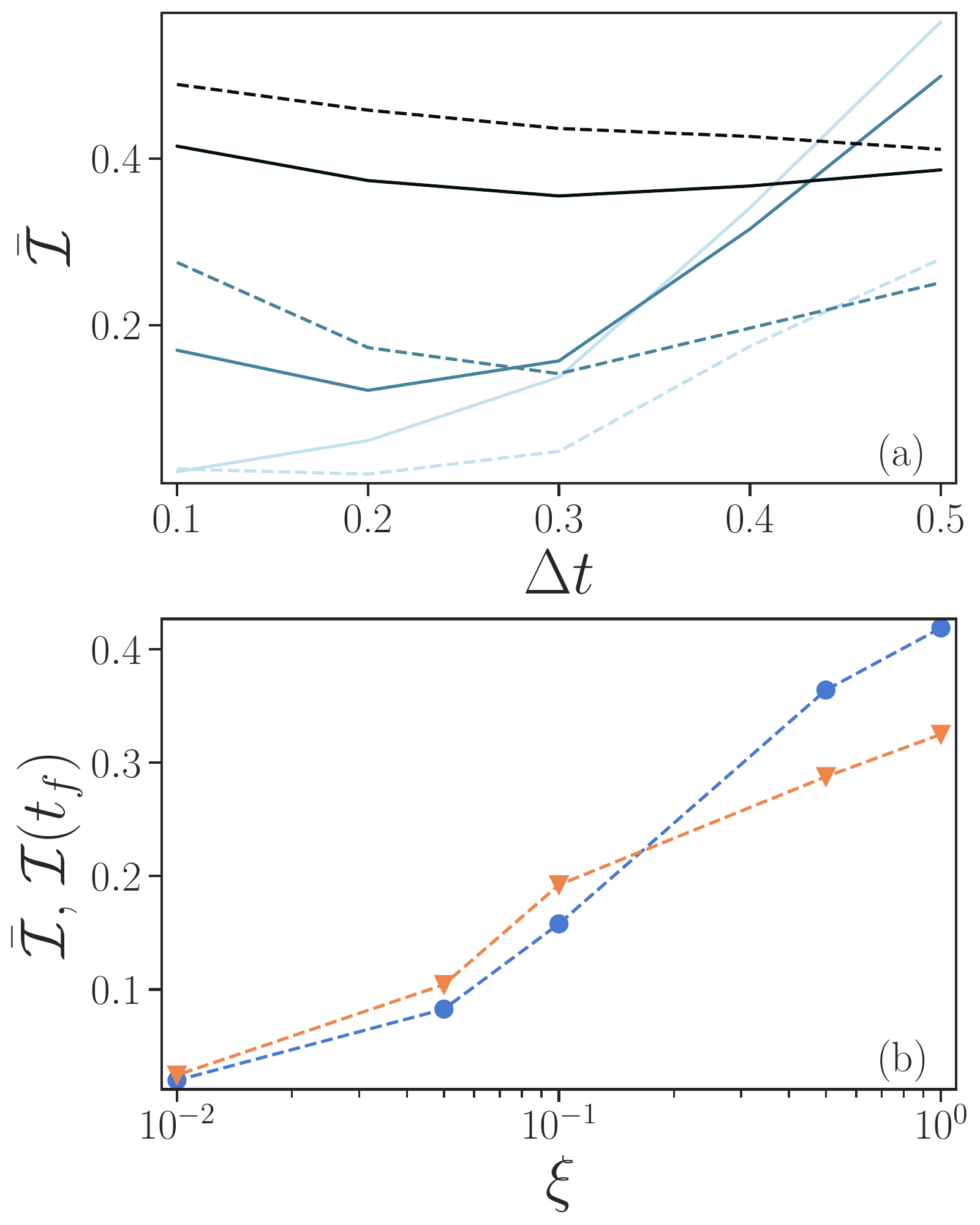}
}
\caption{
\label{fig:noise_fidelity}
(a)
Infidelity averaged over time as a function of time-step size $\Delta t$ for an evolution from $t=0$ to a final time $t_f=2$. Different noise levels $\xi = 0.01,\;0.1,\;1$ are represented by lighter to darker colors. 
(b) Time-averaged (blue circles) and final (orange triangles) infidelity as a function of noise levels. Here the final time is taken as $t_f=2$. and we choose $\Delta t = 0.2$. 
In both panels, results from first-order Trotter implementations are represented by continuous lines, while second-order by dashed lines. 
Parameters \parasf, $\gamma=1$. 
}
\end{figure}

We now turn to more realistic, and thus noisy, devices.
In Fig.~\ref{fig:error_model}, for noiseless simulations, we observed that the infidelity increases monotonously with the time-step size $\Delta t$, and that a second-order Trotterization is always preferred. 
In the presence of noise, however, an increased number of gates can lead to stronger noise effects, and thus instead of improving the quality of the simulations, it may result in worse fidelity. 
In Fig.~\ref{fig:noise_fidelity}(a) we thus consider evolution of the full model, unitary and dissipative part,
up to a time $t=2$ for different magnitudes of noise $\xi = 0.01,\;0.1,\;1$ (from lighter to darker colors) for either a first-order Trotter step (continuous lines) or a second-order Trotter step (dashed lines). In particular, we depict the infidelity versus the length of the time-step $\Delta t$. 
We observe that for intermediate values of noise 
$\xi = 0.1,\;1$
there is an optimal time interval $\Delta t$ that corresponds to the lowest infidelity, and that first-order Trotterization can perform better at smaller $\Delta t$.  

We now consider the open system dynamics case.
The impact of noise on fidelity is depicted in Fig.~\ref{fig:noise_fidelity}(b). Here we show both the average infidelity over the time interval from $t=0$ to $t=2$ (blue line with circles), and the infidelity at the final time (orange line with triangles). 
We consider exclusively a second-order Trotter decomposition and a time-step $\Delta t=0.2$. Fig.~\ref{fig:noise_fidelity}(b) indicates a monotonous growth of infidelity with the noise-factor $\xi$, for the parameters explored. 

In Fig.~\ref{fig:infidelity_over_time} we show the infidelity versus time for first-order (solid lines) and second-order (dashed line) Trotterizations, while $\Delta t=0.2$. We observe that only for small values of $\xi$ one would prefer a second-order Trotterization to improve on the fidelity of the states. We note, not shown here, that for $\xi=0.01$ the dynamics is almost identical to the noiseless case.  



\begin{figure}[h]
\centering
\subfloat{%
\centering
\includegraphics[width=1\columnwidth]{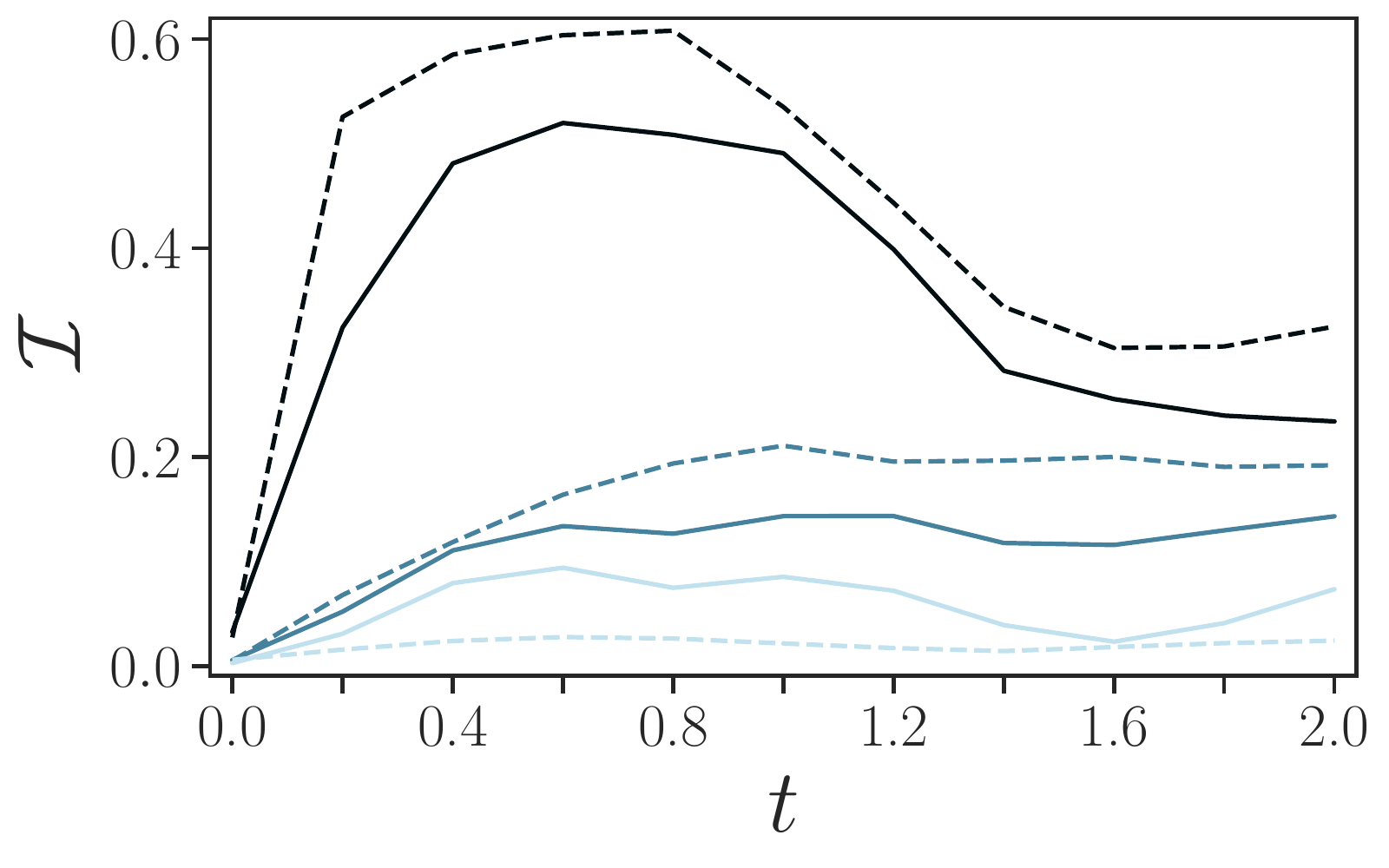}
}
\caption{Infidelity as a function of time in open system simulation in presence of noise.
Using first-order Trotter (solid) and second-order Trotter (dashed) at $\Delta t = 0.2$. 
At noise-factor $\xi = 0.01,\;0.1,\;1$ (from lighter to darker colors).
Other parameters are \parasf, $\gamma=1$.
}
\label{fig:infidelity_over_time}
\end{figure}

To better understand the role of dissipation, we aim to verify its effect on the accuracy of the simulation. 
To focus specifically on the role of $\gamma$, we consider only a second-order Trotter evolution, a fixed value of $\Delta t=0.2$ and $\xi=0.01$, where the simulation of the quantum computer shows generally better performance 
compared to levels of higher magnitudes of noise $\xi=0.1, 1$.
Fig.~\ref{fig:gamma_noise}(a) we plot the time-averaged infidelity at different values of $\gamma$, with (orange line with circles) and without noise (blue line with triangles).
In noiseless simulations the infidelity increases with $\gamma$, while
in noisy simulations the infidelity initially reduces to a minimum at $\gamma=1$.
Our understanding is that the dissipation in the exact calculations acts in a similar way as the intrinsic noise on the device, by drawing the system to its ground state and reducing coherence.
It thus can be easier for a lossy quantum hardware to simulate a lossy system compared to a closed system ($\gamma=0$). However, a system with larger $\gamma$ also implies further difficulties in the simulations stemming, for example, from Trotterization.
It thus occurs that the intrinsic dissipative dynamics can, in some regimes, be better represented on a noisy device.

In Fig.~\ref{fig:gamma_noise}(b) we plot the infidelity versus time for different values of the dissipative rate $\gamma$.
We observe that for 
$\gamma \le 1$
the infidelity tends to increase with time, while for larger values of $\gamma\ge 1.5$, the infidelity can decrease after 
a maximum at an earlier time $t\approx0.4$.



\begin{figure}[h]
\centering
\subfloat{%
\centering
\includegraphics[width=1\columnwidth]{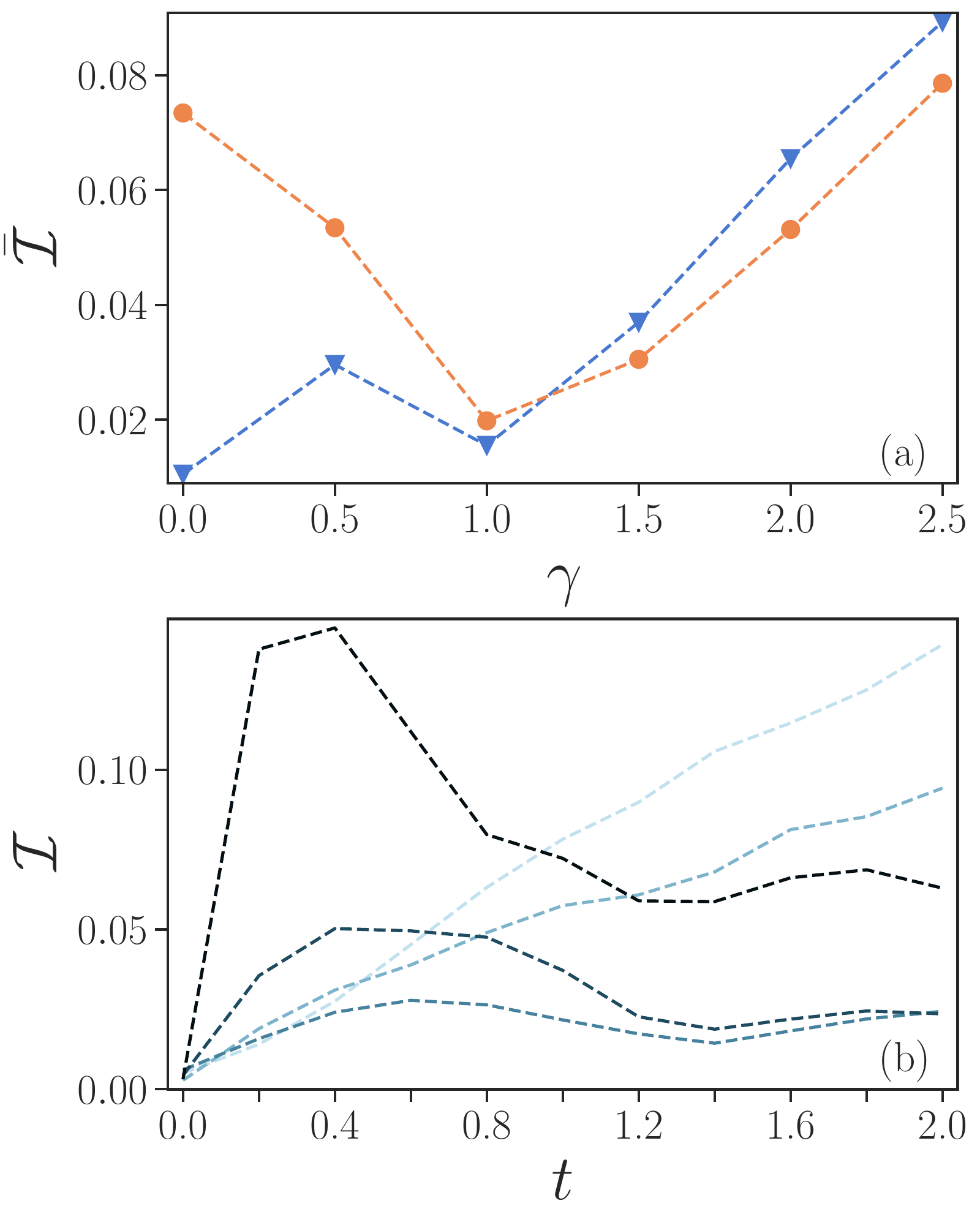}
}
\caption{(a) Infidelity averaged over time versus dissipative rate $\gamma$ with noise $\xi=0.01$ (orange line with circles) and without noise (blue line with triangles).
(b) Infidelity as a function of time $\gamma = 0,\;0.5,\;1,\;1.5,\;2,\;2.5$ (from lighter to darker colors).
Second order Trotter at $\Delta t = 0.2$ and the other parameters are \parasf.
}
\label{fig:gamma_noise}
\end{figure}







Fig.~\ref{fig:expval_1f} shows the average occupation in the harmonic oscillator, panel (a), and the expectation values of $\hat{\sigma}^z$ of the spin, panel (b), versus time. 
In both panels the dotted line corresponds to the exact values, 
 solid and dashed lines to $\xi=0.01,\;0.1,\;1$, respectively from lighter to darker shades, and solid lines are used for first-order Trotterizations, while dashed lines for second-order. For each noise level $\xi$ we have used the Trotterization order which corresponds to the lower fidelity. 

The oscillatory evolution of the occupation of the harmonic oscillator is captured, only partially, with the smaller non-zero noise parameter considered $\xi=0.01$, panel (a), while the occupation of the harmonic oscillator at $\xi=1$ quickly stagnates at around a value of $1$. 
%
Instead, the simpler evolution of $\hat{\sigma}^z$ is captured fairly well also for the different values of $\xi$, as the simulated dissipation of the spin is closer to the relaxation of the spin-qubit under noise.  

\begin{figure}[h]
\centering
\subfloat{%
\centering
\includegraphics[width=1\columnwidth]{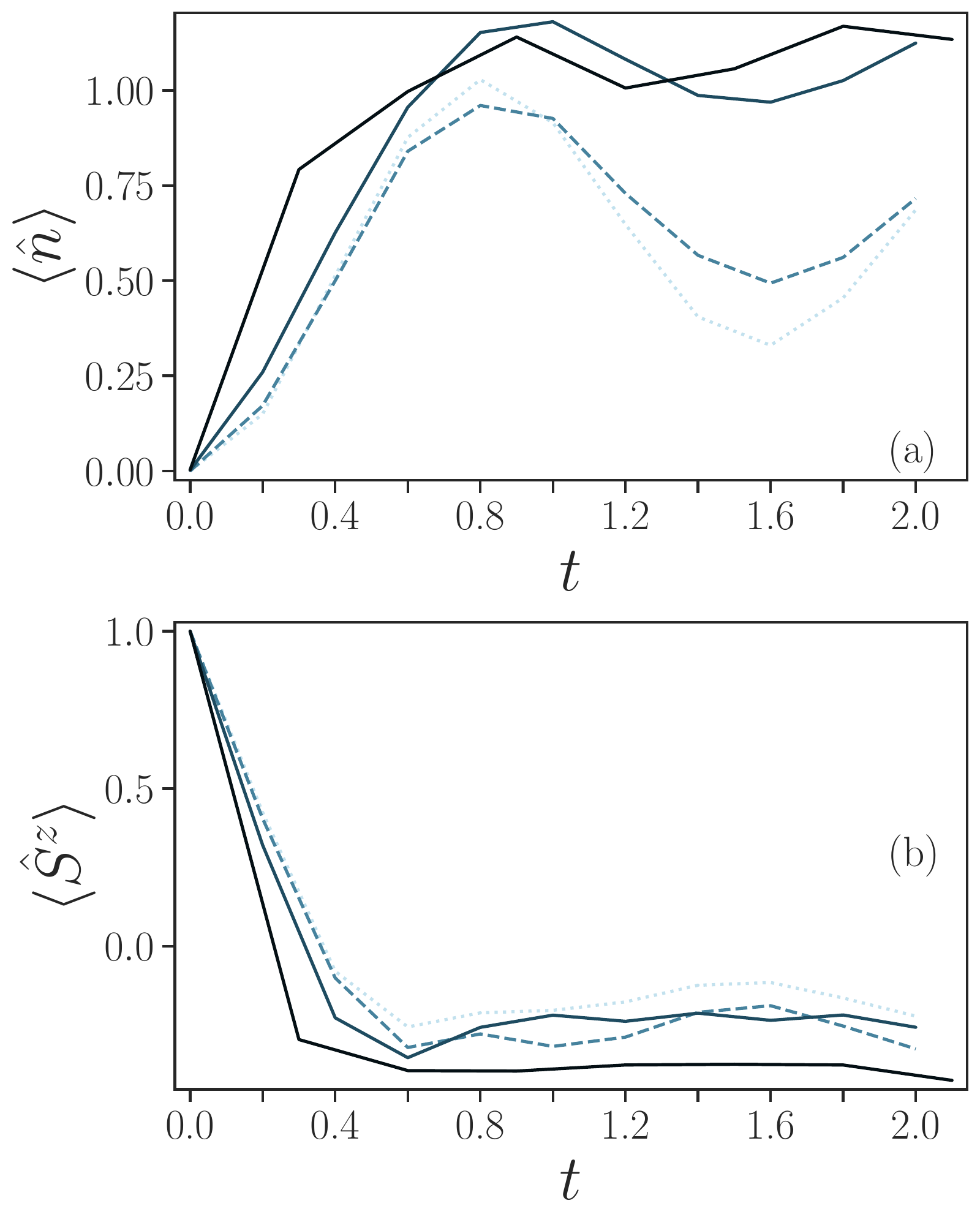}
}
\caption{
(a) Average bosonic occupation $\expect{\hat{n}} $ and (b) $\expect{\sz} $ as function of time. 
Different noise levels $\xi = 0.01,\;0.1,\;1$ are presented, respectively by lighter to darker colors. As a reference, exact simulations are depicted by dotted lines. Results obtained using first-order Trotterization are with solid lines, while second-order with dashed lines.  
Other parameters are \parasf  ~and $\gamma=1$.
}
\label{fig:expval_1f}
\end{figure}

%
\subsection{Two-spin system}\label{sec:twospins} 

\begin{figure}[h]
\centering
\subfloat{%
\centering
\includegraphics[width=1\columnwidth]{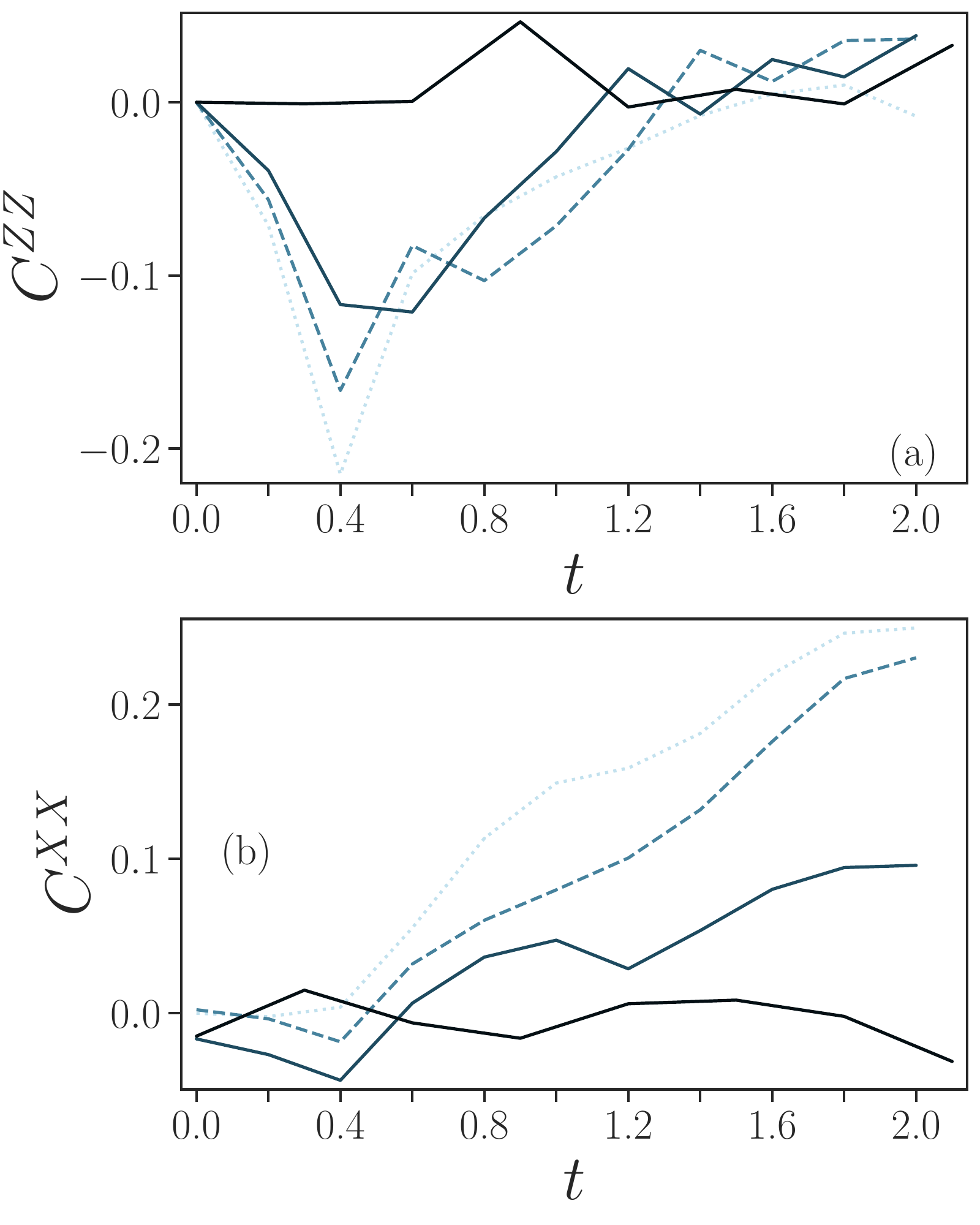}
}
\caption{
Correlations for the case of two spins. 
(a) spin-z connected correlation $\szcorr$ 
(b) spin-x connected correlation  $\sxcorr$ as a function of time. 
Different noise levels $\xi = 0.01,\;0.1,\;1$ are presented, respectively by lighter to darker colors. 
As a reference, exact simulations are depicted by the dotted lines. Results obtained using first-order Trotterization are with solid lines, while second-order with dashed lines.  
Other parameters are \parasff ~and $\gamma=1$. 
}
\label{fig:expval_2f}
\end{figure}

We here extend the system to two spins to see whether it is possible to study correlation developing between them through a mediated interaction via the harmonic oscillator, as the two spins do not directly interact with each other.   
We use the parameters \parasff \ and $\gamma=1$.
We prepare the initial state in a product state of one spin in the excited state, one in the ground state, and the harmonic oscillator completely empty. This can allow us to observe non-trivial dynamics while still requiring just a few occupied level of the harmonic oscillator. 

As for the single spin simulations, we first evaluate the infidelity in the presence of noise.
Simulating two spins requires roughly twice the number of gates as simulating one spin.
A single $\Delta t$ evolution with a first-order Trotter requires $113$ single-qubits and $36$ \cxgate s, while the second-order Trotter requires $177$ single-qubits and $70$ CX gates, see Fig.~\ref{fig:ibmq_jakarta} and Table \ref{tab:encoding_depth} in Appendix \ref{sec:boson_to_qubit}.  
Also in the case of two spins, we find that the optimal Trotter time-step $\Delta t$ to be the same as for the single spin case (not shown).

To study the emerging correlations 
between the spins mediated by interaction with the photons,
we consider the spin-spin correlators  
\begin{align}
    \szcorr = \szcorrdef \nonumber \\ 
    \sxcorr = \sxcorrdef.  
\end{align}
These connected correlation functions (also called second-order Ursell functions or cumulants) corresponds to the covariance in statistics and vanish if and only if $\hat{\sigma}^{\left(\cdot \right)}_1$ and $ \hat{\sigma}^{\left(\cdot \right)}_2$ are statistically independent \cite{Ursell1927, Percus1975, Shlosman1986}.

In Fig.~\ref{fig:expval_2f} we show $\szcorr$ and $\sxcorr$ for, again, $\xi=0.01,\;0.1,\;1$ from lighter to darker lines. The solid lines correspond to first-order Trotter and dashed lines to second-order Trotter and these Trotterization orders have been chosen as they result, for the respective amount of noise, to the lowest infidelity. 
In both panels the dotted lines correspond to the exact values.  
The exact case simulations show a build-up in anti-correlation in $z$-direction at $t=0.4$, before reducing to $0$ which can be observed already for $\xi=0.1$. A correlation in $x$-direction builds up monotonously over time and one would need $\xi=0.01$ for a clearer signal. 

In principle, correlations could be observed for higher number of spins.
In practice, the larger number of qubits needed, and their connectivity, would result in an increased number of gates which would limit the fidelity in NISQ devices. 
We also note that going from one to two spins we had to increase $\omega$ to keep the higher levels of the harmonic oscillator sparsely populated. If one does not want to increase the number of levels studied for the harmonic oscillator, a similar adjustment, like decreasing the coupling between the harmonic oscillator and the spins, would be necessary when increasing the number of spins. 


\section{Conclusions}
In this paper, we have studied the feasibility of simulating open spin-boson dynamics on a quantum computer. 
We used a second-quantization mapping of the bosonic degrees of freedom and Trotterization of the unitary to implement the Hamiltonian.
To implement the dissipative dynamics, we used collisions and resets with auxiliary qubits. 

We found that in our parameter regime, the Hamiltonian simulation is the limiting factor to the fidelity.
We surveyed optimal Trotterization formulas and time-step sizes depending on the level of noise in the system. 
We selected the open dissipative rate with the highest fidelity in noisy circuits, and we found that current noise levels in the machine we considered would make such simulations particularly challenging.

Anticipating future improved devices, we ran our simulations on 10\% and 1\% of current noise levels, and we were able to show that it would be possible to attain much higher fidelities. Furthermore, certain observables could be well represented with larger amounts of noise. 
Importantly, the simulation of an open system can be more accurate than unitary evolution as the open system dynamics could be closer to how a noisy computer is already affecting a state.

Future developments in noise reduction in the hardware, in post-processing error mitigation and also in reducing the number of gates for unitary evolutions can lead to significant increase in simulation power. 

In our system we have limited the dissipation to the spins. 
An interesting avenue for future work could be the inclusion of loss in the bosonic degrees of freedom of the cavity, for which additional auxiliary qubits, gates and connectivity requirements could prove challenging.


{\it Acknowledgement}: AB acknowledges support from Ministry of Education 
 of Singapore AcRF MOE Tier-II (Project No. T2MOE2002). 
KLC and DP acknowledge support from the National Research Foundation, Singapore under its QEP2.0 programme (NRF2021-QEP2-02-P03).

\FloatBarrier
\appendix

\renewcommand\thefigure{\thesection.\arabic{figure}}    
\setcounter{figure}{0}   

\section{Encoding of bosonic operators onto qubits \label{sec:boson_to_qubit}}

We will quickly review the d-level-to-qubit mapping we used to encode the bosonic operators as strings of Pauli matrices.
The method and different binary encodings are discussed in \cite{Sawaya2020}. The steps can be summarized as:
\begin{enumerate}
    \item Truncate the infinite-dimensional harmonic oscillator at some level $\dho$  
    \item Rewrite each bosonic operator $\hat{A}$ as a sum of level transitions 
    \begin{align} 
    \nonumber
    \hat{A} = \sum^{\dho-1}_{l,l'=0} a_{l,l'}\ket{l}\bra{l'}, \ \hat{A} = \{ \opa, \opad, \ada\}
    \end{align}
    \item Assign each level an integer $\ket{l} \xrightarrow{\rm integer} \ket{i},\ i \in \mathbb{N}$ 
    \item Write each integer in binary 
    \begin{align}
    \nonumber
    \ket{i} \xrightarrow{\rm integer-to-bit} \bigotimes^{Q_B}_{m=1} \ket{b_m},\ b_m \in \{0,1\}
    \end{align}
    \item Map each bit pair $\ket{b_m} \bra{b'_m}$ to Pauli matrices using 
    \begin{align} \label{eq:bittopaulis}
        \ket{0}\bra{0} &= \frac{1}{2}\left( \ident + \pz\right) \nonumber \\
        \ket{1}\bra{1} &= \frac{1}{2}\left( \ident - \pz\right) \nonumber \\
        \ket{0}\bra{1} &= \frac{1}{2}\left( \px + i\py\right) = \ppl \nonumber \\
        \ket{1}\bra{0} &= \frac{1}{2}\left( \px - i\py\right) = \pmi \nonumber
    \end{align}
\end{enumerate}

The result is that each level transition is written as a string of Pauli operators and each bosonic operator $\hat{A}$ as a sum of $N_P$ Pauli strings
\begin{equation}
    \hat{A} = 
    \sum^{N_P}_{k=1} c_{k} \bigotimes^{Q_B}_{j=1}\hat{\sigma}_{kj},\ 
    \hat{\sigma}_{kj} \in \{\ident, \px, \py, \pz\}
\end{equation}
Where $ Q_B = \lceil \sqrt{\dho}\rceil$ is the number of qubits which encode the bosonic levels ($ \lceil \cdot\rceil$ is the ceiling function).

\paragraph{Gate Requirements}
When writing the integers in binary in step 4, different integer-to-bit encodings result in different Pauli strings and ultimately in a different representation of the Hamiltonian.
While the representations of the Hamiltonian are theoretically equivalent, they come with different gate counts and thus result in different performances on noise devices.

As integer-to-bit encodings we considered Standard Binary and Gray code, since both of them are compact, i.e. require the minimum amount of qubits. 
Table \ref{tab:encoding_depth} shows the gates required to evolve one time-step of the trotterized unitary $e^{-i \hsb \dt}$ and dissipation on the ibmq\_jakarta device.
This includes additional \cxgate s to implement any necessary SWAP-Gates due to limited qubit connectivity (Fig.~\ref{fig:ibmq_jakarta}).
For our Hamiltonian $\hsb$ Gray Code yielded less gates than Standard Binary in all cases, which is why we used Gray Code throughout the main text.


\begin{figure}
\centering
\includegraphics[width=0.4\columnwidth]{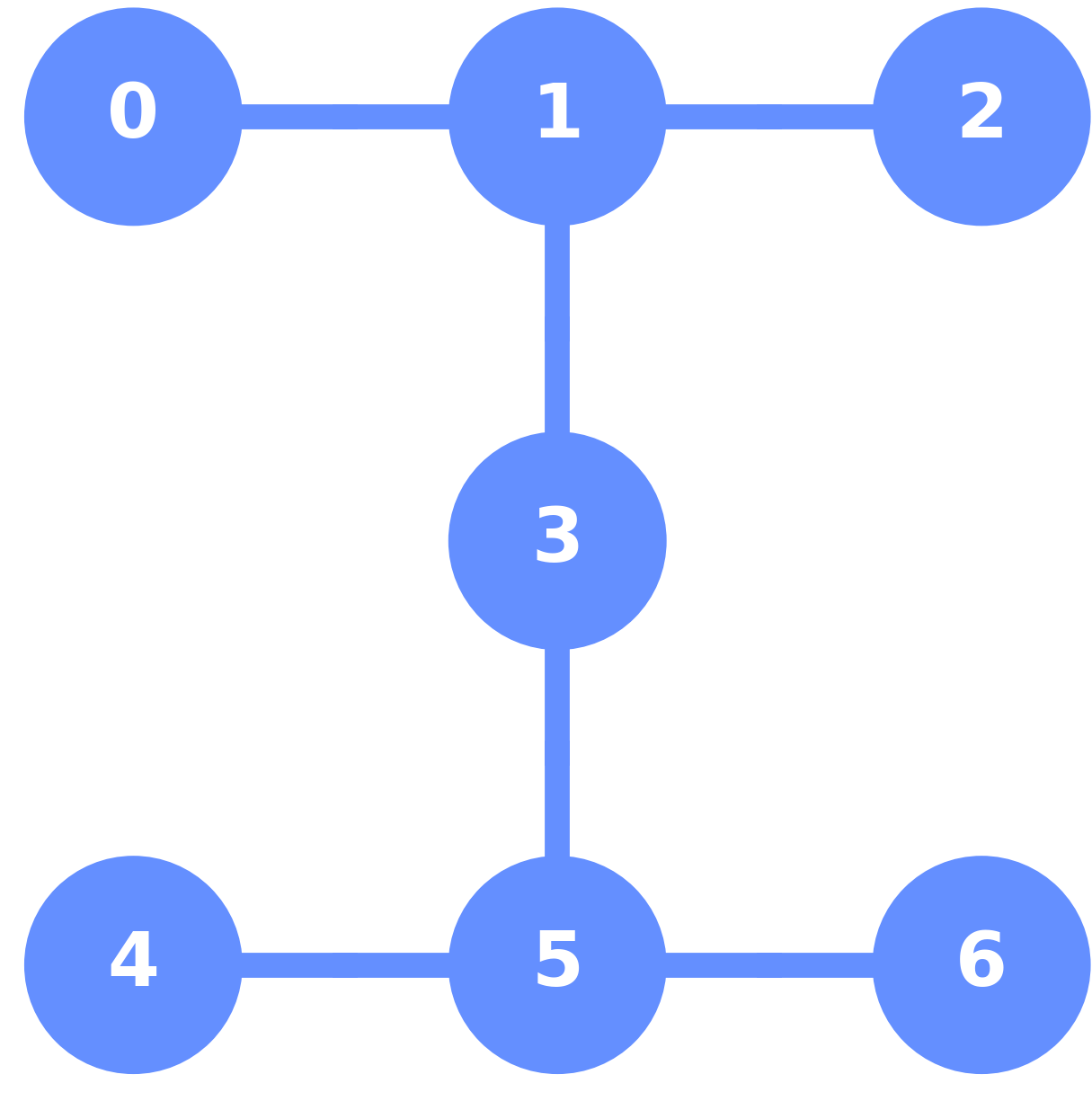}
\caption{
Qubit connectivity of the used ibmq\_jakarta device
\label{fig:ibmq_jakarta}
}
\end{figure}

\begin{table}
\centering
\subfloat{
\resizebox{.9\columnwidth}{!}{
\begin{tabular}{ccc|cc|cc|}
\cline{4-7}
                         &                          &               & \multicolumn{2}{c|}{Standard Binary} & \multicolumn{2}{c|}{Gray Code}    \\ \hline
\multicolumn{1}{|c|}{$\numspins$} & \multicolumn{1}{c|}{$\dho$} & Trotter order & \multicolumn{1}{c|}{Single}   & CX   & \multicolumn{1}{c|}{Single} & CX  \\ \hline
\multicolumn{1}{|c|}{1}  & \multicolumn{1}{c|}{4}   & first         & \multicolumn{1}{c|}{53}       & 21   & \multicolumn{1}{c|}{94}     & 43  \\ \hline
\multicolumn{1}{|c|}{1}  & \multicolumn{1}{c|}{4}   & second        & \multicolumn{1}{c|}{94}       & 34   & \multicolumn{1}{c|}{75}     & 28  \\ \hline
\multicolumn{1}{|c|}{1}  & \multicolumn{1}{c|}{8}   & first         & \multicolumn{1}{c|}{156}      & 66   & \multicolumn{1}{c|}{122}    & 60  \\ \hline
\multicolumn{1}{|c|}{1}  & \multicolumn{1}{c|}{8}   & second        & \multicolumn{1}{c|}{282}      & 124  & \multicolumn{1}{c|}{191}    & 107 \\ \hline
\multicolumn{1}{|c|}{2}  & \multicolumn{1}{c|}{4}   & first         & \multicolumn{1}{c|}{106}      & 37   & \multicolumn{1}{c|}{122}    & 36  \\ \hline
\multicolumn{1}{|c|}{2}  & \multicolumn{1}{c|}{4}   & second        & \multicolumn{1}{c|}{191}      & 65   & \multicolumn{1}{c|}{168}    & 74  \\ \hline
\multicolumn{1}{|c|}{2}  & \multicolumn{1}{c|}{8}   & first         & \multicolumn{1}{c|}{270}      & 139  & \multicolumn{1}{c|}{200}    & 156 \\ \hline
\multicolumn{1}{|c|}{2}  & \multicolumn{1}{c|}{8}   & second        & \multicolumn{1}{c|}{496}      & 272  & \multicolumn{1}{c|}{409}    & 255 \\ \hline
\end{tabular}
} 
}
\caption{
Gate counts, both for \cxgate{} and single-qubit gates, to evolve one time-step of master equation (\ref{eq:gksl}) on the jakarta device. 
}
\label{tab:encoding_depth}
\end{table}

\paragraph{Mapped Hamiltonian}
After the mapping of the harmonic oscillator to qubits, the Hamiltonian (Eq.~\ref{eq:spinboson}) is written as a sum of Pauli strings  $h_{\sumindex}$.
The unitary $e^{\sum_{\sumindex} h_{\sumindex}}$ is then trotterized (Eq.~\ref{eq:firstorder}).
The mapped Hamiltonian $\hsb=\sum_{\sumindex} h_{\sumindex}$ we implemented for the main text reads explicitly 
\begin{align}
\hsb =& - \sqrt{2} {\sigma^{x}}_{0} {\sigma^{x}}_{1} {\sigma^{z}}_{2} + \sqrt{2} {\sigma^{x}}_{0} {\sigma^{x}}_{1} \nonumber
\\\nonumber &+ (1-\sqrt{3}) {\sigma^{x}}_{0} {\sigma^{x}}_{2} {\sigma^{z}}_{1} + (1+\sqrt{3}) {\sigma^{x}}_{0} {\sigma^{x}}_{2} 
\\ &+ \frac{1}{4}{\sigma^{x}}_{0} 
- \frac{1}{2} {\sigma^{z}}_{0} - 2 {\sigma^{z}}_{1} {\sigma^{z}}_{2} - 4 {\sigma^{z}}_{1}  \label{eq:hsb_digit_1} 
\end{align}
for the single spin case, and
\begin{align} 
\hsb =& - \sqrt{2} {\sigma^{x}}_{0} {\sigma^{x}}_{1} {\sigma^{z}}_{2} + \sqrt{2} {\sigma^{x}}_{0} {\sigma^{x}}_{1} \nonumber
\\\nonumber &+ (1-\sqrt{3}) {\sigma^{x}}_{0} {\sigma^{x}}_{2} {\sigma^{z}}_{1} + (1+\sqrt{3}) {\sigma^{x}}_{0} {\sigma^{x}}_{2} 
\\\nonumber &+ \frac{1}{4}{\sigma^{x}}_{0} 
- \sqrt{2} {\sigma^{x}}_{1} {\sigma^{x}}_{3} {\sigma^{z}}_{2} + \sqrt{2} {\sigma^{x}}_{1} {\sigma^{x}}_{3} 
\\\nonumber & + (1-\sqrt{3}) {\sigma^{x}}_{2} {\sigma^{x}}_{3} {\sigma^{z}}_{1} + (1+\sqrt{3}) {\sigma^{x}}_{2} {\sigma^{x}}_{3} 
\\\nonumber &+ \frac{1}{4}{\sigma^{x}}_{3} 
- \frac{1}{2} {\sigma^{z}}_{0} - 3 {\sigma^{z}}_{1} {\sigma^{z}}_{2} - 6 {\sigma^{z}}_{1} \\ &- \frac{1}{2} {\sigma^{z}}_{3}  \label{eq:hsb_digit_2}
\end{align}
for two spins case. Each term constitutes one of the $h_{\sumindex}$ in Eqs.~(\ref{eq:firstorder},\ref{eq:secondorder}).

\section{Noise model\label{sec:noise_model}}

Qiskit supplies noise models based on device properties measured during calibration. 
In order to simulate improved future device, we engineer our noise from an identical model, but from lower noise levels.

The noise model contains three error sources \cite{Georgopoulos2021}
(i) thermal relaxation (relaxation and dephasing)
(ii) depolarizing (Pauli) error 
(iii) readout (measurement) error.
At every gate, first the thermal relaxation and then the depolarizing error is applied.
The strength of the depolarizing error is calculated backwards, to reach a target 'gate error' when combined with the thermal relaxation. Details can be found at \cite{qiskit_doc}.

\subsection{Error Sources}
\subsubsection{Thermal Relaxation Error}
Thermal relaxation is defined by the qubit-specific parameters $\tOne$ time, $\tTwo$ time, qubit frequency $\freqQubit$ and qubit temperature $\tempQubit$.
The thermal error channel is then given time to act according to a gate-dependent gate time.
For two-qubit-gates, the error is simply the tensor product between two single-qubit channels.

$\tOne$ is qubit-specific time until relaxation, i.e. to decay from the excited state to the ground state.
$\tTwo$ qubit-specific coherence time, or time until dephasing.
The qubit frequency $\freqQubit$ is the difference in energy between the ground and excited states.
The qubit temperature $\tempQubit$ is assumed to be $0$ in Qiskit's and our noise models.

The qubit frequency and temperature enter only via the excited state population.
If $\freqQubit \rightarrow \infty$ or $\tempQubit=0$, the excited state population is $0$.
Since $\tempQubit=0$ in our models, both the frequency and temperature can effectively be ignored as parameters.



For $T_2 < T_1$, thermal relaxation is most straight-forwardly described by (assuming the device to be at $0$ temperature) 
\begin{align}
K_{T_0} &= \sqrt{\probp_I} \ident
,\\ 
K_{T_1} = \sqrt{\probp_Z} \pz
,&\ 
K_{T_2} =\sqrt{\probp_{reset}} \ket{\downarrow} \bra{\downarrow} \\
\channelthermal(\dm) &= \sum_{i=10}^2 K_{T_{\sumindex}} \dm K^\dagger_{T_{\sumindex}}
\end{align}
It is composed of the probabilities of a phase-flip $\probp_Z$, a reset to the ground state $\probp_{reset}$, or for nothing to happen $ \probp_{\ident}$.
The probabilities $\probp_Z$, $\probp_{reset}$ are calculated of $T_1$, $T_2$ and the gate time $\timeGate$.
\begin{align}
\probp_{reset} =& 1 - \probp_{T_1} = 1 - e^{-\timeGate \mult \frac{1}{T_1}}\\
\probp_Z =& (1 - \probp_{reset})\left(1 - \frac{\probp_{T_2}}{ \probp_{T_1} }\right)/2 \\
=& (1 - \probp_{reset}) \left(1 - e^{-\timeGate \mult (\frac{1}{T_2} - \frac{1}{T_1})}\right) /2 
\label{eq:prob_thermal_dephasing}
\\
\probp_{\ident} =& 1 - \probp_Z - \probp_{reset}. 
\end{align}

If $2T_1 \ge T_2 > T_1$ thermal relaxation has to be described by it's Choi matrix
\begin{align}
\dm \rightarrow \channelthermal (\dm) = tr_1 [C(\dm^T \otimes I)] \\
C_{\channelthermal} = 
\left(
\begin{matrix}
1 & 0 & 0 & \probp_{T_2} \\
0 & 0 & 0 &  0 \\
0 & 0 & \probp_{reset} & 0 \\
\probp_{T_2} & 0 & 0 & 1-\probp_{reset}
\end{matrix}
\right)
\end{align}
Which can also be used if $T_2 < T_1$ to compute the process fidelity in Eq.~(\ref{eq:process_fidelity}).

At the time of writing all qubits on the Jakarta hardware satisfied $T_2 < T_1$. 
This is not necessarily the case for all devices provided by IBM or in general.

\subsubsection{Depolarizing Error}
The depolarizing noise (or Pauli) channel is composed of either a bit-flip ($\px$), a
phase-flip ($\pz$) or both at the same time ($\py$), all with equal probability \cite{Georgopoulos2021}.
\begin{align}
\dm \rightarrow
\channeldepol(\dm) &= \sum_{i=1}^3 K_{\probp_{\sumindex}} \dm K^\dagger_{\probp_{\sumindex}} 
\\
K_{\probp_0} = \sqrt{1-\probp_D} \ident
&,\ 
K_{\probp_1} = \sqrt{\frac{\probp_D}{3}} \px
\\
K_{\probp_2}  = \sqrt{\frac{\probp_D}{3}} \py
&,\
K_{\probp_3} = \sqrt{\frac{\probp_D}{3}} \pz
\end{align}

\paragraph{Gate Infidelity} 
The probability of a depolarizing error is calculated from the target gate infidelity $\ifidGate$, and the infidelity due to thermal relaxation $\ifidThermal$. 
\begin{align}
    \label{eq:ifiddepol}
    \ifidDepol = \ifidGate - \ifidThermal
\end{align}
The target gate infidelity is given as a parameter, while $\ifidThermal$ has to be calculated as
\begin{align}
\label{eq:ifidthermal}
\fid_{T} &= 1 - \ifidThermal \\
   &= \fid_{avg}(\channelthermal, U) \\
    &= \int d\psi \langle\psi|U^\dagger
        \channelthermal(|\psi\rangle\!\langle\psi|)U|\psi\rangle \\
    &= \frac{d \ifid_{pro}(\channelthermal, U) + 1}{d + 1}
\end{align}
where $\ifid_{pro}(\channelthermal, U)$ is the
process fidelity of the input quantum channel $\channelthermal$ with a target unitary $U$, and
$d$ is the dimension of the channel.

\begin{align}
\label{eq:process_fidelity}
    \ifid_{pro}(\channelthermal, \mathcal{F})
            = F(C_{\channelthermal} / d, \rho_{\mathcal{F}})
\end{align}
where $\fid$ is the state fidelity as defined in the main text
\begin{align}
    \fid(\rho_1, \rho_2) = \left( \text{Tr} \left[ \sqrt{\sqrt{\rho_1}\rho_2\sqrt{\rho_1}} \right] \right)^2
\end{align}
$C_{\channelthermal} / d$ is the
normalized Choi matrix for the channel
$\channelthermal$, and $d$ is the input dimension of
$\channelthermal$.

Importantly for our reduced-noise models, the infidelity from thermal relaxation $\ifidThermal$ is linear in the gate time $\timeGate$.
Thus, when we rescale $\ifidGate \rightarrow \xi \mult \ifidGate$, $\timeGate \rightarrow \xi \mult \timeGate$, we indirectly scale $\ifidDepol \rightarrow \xi \mult \ifidDepol$, $\ifidThermal \rightarrow \xi \mult \ifidThermal$.
This way the relative contribution of the error channels $\ifidDepol/\ifidThermal$ to the infidelity remains unchanged.

\paragraph{Depolarizing Error Probability}
If we write the depolarizing error in terms of the identity and the complete depolarizing channel $D$, we can rewrite the gate fidelity
\begin{align}
\mathcal{E}_{D} &= (1-\probp_D) \mult \ident + \probp_D \mult D \\
	\fid_{gate} &= 1 - \ifidGate \\
    &= \fid(\channeldepol \mult \channelthermal) \\
    &= (1-\probp_D) \fid_T + \probp_D\mult\fid_D\\
    &= \fid_T  - \probp_D\mult(d\mult\fid_T - 1) / d
\end{align}
Where $d= 2^{qubits}$ is the dimensionality of the gate.
From this the solution for the depolarizing error probability is
\begin{align}
	\probp_D &= d(\fid_T - \fid_{gate}) / (d\mult\fid_T - 1) \\
    &= d(\ifidGate - \ifid_T) / (d\mult\fid_T - 1) 
\end{align}
More details can be found at \cite{qiskit_doc}. 

\subsubsection{Measurement Error}
A measurement error is equivalent to a bit-flip $\px$ followed by a noiseless readout \cite{Georgopoulos2021}.
The probability of the readout error $\probp_R$ is given by the probability $P(n|m)$ of recording a noisy measurement outcome as $n$, given the true measurement outcome is $m$.
\begin{align}
  	K_{R_0}  = &\sqrt{1-\probp_R} \ident
  ,\ 
  K_{R_1} = \sqrt{\probp_R} \px
  \\
  &\probp_R = \sum_{n \neq m} P(n|m)
\end{align}
Where $n$, $m$ run over all qubits, in the case of two qubits $n,m \in \{00, 01, 10, 11\}$.
See \cite{qiskit_doc} for further details. 

\subsubsection{Error Sources in Reference Device}
Given the three error sources, one can ask which error source causes the dominant contribution to the noise in our results. 
As we use measurement error mitigation and it is independent of the circuit depth, we will ignore the measurement error.
Instead we focus on the ratio of the thermal and depolarizing errors in contributing to the infidelity, $\ifidThermal/\ifidDepol$.
To give a rough estimation, we assume all gates $ g \in \{CNOT, RZ, SX, X\}$ and all qubits $q$ are used equally often, and average over both.
\begin{align}
    \ifidThermal/\ifidDepol = 
    \frac{1}{N_q} \sum_{q=1}^{N_q=7} \left( 
    \frac{1}{N_g} \sum_{g}^{N_g=4} \left(
    \frac{\ifidThermal \left( q, g \right)}{\ifidDepol \left( q, g \right)}
    \right)
    \right)
\end{align}
We calculate $\ifidThermal \left( q, g \right)$ and $\ifidDepol \left( q, g \right)$ using Eqs.~(\ref{eq:ifidthermal}) and (\ref{eq:ifiddepol}) respectively, and get the current calibration data from IBM.
At the time of writing, the result for the Jakarta device is $\ifidThermal/\ifidDepol=15.4$. 
We conclude that thermal relaxation is the main source of infidelity in our simulations, by one order of magnitude compared to depolarization.

\subsubsection{Calibration Data}
We base our reduced-noise models on the same hardware that we run our full-noise circuits on, the 7 qubit IBMQ Jakarta device. 

At the time of writing the calibration data is: \\
Processor: Falcon r5.11H, V1.1.0 \\
Avg. \cxgate{} Error: $1.109e^{-2} $ \\
Avg. Readout Error: $3.349e^{-2} $ \\
Avg. $\tOne$: $139.01$ us \\
Avg. $\tTwo$: $44.82$ us \\
Avg. Gate time: $454.095$ ns \\
Avg. Qubit Frequency: $5.08$ GHz \\
Avg. Qubit Anharmonicity $-0.329$ GHz \\
For more details see \cite{qiskit_resources}.

\setcounter{figure}{0}

\section{Gate Definition}\label{sec:gate_definition}
Some of the gates used are defined here.
A controlled operation $CO$ is defined as 
\begin{align}
\label{eq:def_ry}
CO(\theta) &=
I \otimes |\downarrow \rangle\langle \downarrow | + O(\theta) \otimes |\uparrow \rangle\langle \uparrow|,  
\end{align}
where the operation $O$ is a $X$-gate in case of the \cxgate{}, or a rotation around the y-axis $R_Y$ or z-axis $R_Z$.
$R_Y$ and $R_Z$ are respectively defined as
\begin{align}
    R_Y(\theta) &= \exp\left(-i \frac{\theta}{2} Y\right) =
    \begin{pmatrix}
        \cos{\frac{\theta}{2}} & -\sin{\frac{\theta}{2}} \\
        \sin{\frac{\theta}{2}} & \cos{\frac{\theta}{2}}
    \end{pmatrix},
    \\
    R_Z(\lambda) &= \exp\left(-i\frac{\lambda}{2}Z\right) =
    \begin{pmatrix}
        e^{-i\frac{\lambda}{2}} & 0 \\
        0 & e^{i\frac{\lambda}{2}}
    \end{pmatrix}. 
\end{align}

Furthermore, the $\sqrt{X}$-gate is given by 
\begin{align}
    \sqrt{X} = \frac{1}{2} \begin{pmatrix}
        1 + i & 1 - i \\
        1 - i & 1 + i
    \end{pmatrix}. 
\end{align}


\section{Transpiled Circuits \label{sec:transpiled} }

The amplitude damping circuit as in Fig.~\ref{fig:adc_circuit} uses gates which are not available on the quantum computer we were using. 
Instead the IBM Jakarta device uses the gate set \{CNOT, ID, RZ, SX, X\}.
The amplitude damping circuit, in terms of these gates and as it was implemented on the hardware, is in Fig.~\ref{fig:circuits_dissipation}.

\begin{figure}
\centering
\newcommand{\genericry}{ } 
\subfloat{%
\scalebox{1}{
\Qcircuit @C=0.8em @R=0.3em @!R { \\
\nghost{{s} :  } & \lstick{{s} :  } & \gate{\mathrm{R_Z}\genericry} & \gate{\mathrm{\sqrt{X}}} & \qw & \qw & \ctrl{1} & \gate{\mathrm{\sqrt{X}}} & \qw & \qw & \qw\\
\nghost{{a} :  } & \lstick{{a} :  } & \gate{\mathrm{\sqrt{X}}} & \gate{\mathrm{R_Z}\genericry} & \gate{\mathrm{\sqrt{X}}} & \gate{\mathrm{R_Z}\genericry} & \targ & \gate{\mathrm{R_Z}\genericry} & \qw & \qw & \qw\\
\\
\nghost{{s} :  } & \lstick{{s} :  } & \gate{\mathrm{R_Z}\genericry} & \gate{\mathrm{\sqrt{X}}} & \ctrl{1} & \gate{\mathrm{R_Z}\genericry} & \gate{\mathrm{\sqrt{X}}}
& \gate{\mathrm{R_Z}\genericry} & \qw & \qw & \qw\\
\nghost{{a} :  } & \lstick{{a} :  } & \qw & \qw & \targ & \gate{\mathrm{R_Z}\genericry} & \gate{\mathrm{X}} & \qw & \gate{\mathrm{\ket{\downarrow}}} & \qw & \qw\\
\\ }
} 
} 
\caption{
\label{fig:circuits_dissipation}
The dissipation circuit represented in Fig.~\ref{fig:adc_circuit} in terms of the gates available on the IBM Jakarta device. Both lines for qubits $s$, $a$ continue from the first row to the second.
}
\end{figure}
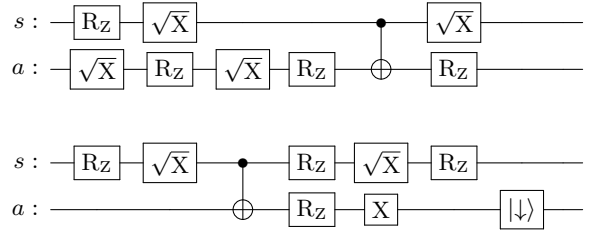


\bibliographystyle{apsrev4-2}
\bibliography{main}

\end{document}